\documentclass[10pt,letterpaper]{article}
\usepackage[top=0.85in,left=2.75in,footskip=0.75in]{geometry}

\usepackage{changepage}

\usepackage[utf8]{inputenc}

\usepackage{textcomp,marvosym}

\usepackage{fixltx2e}

\usepackage{amsmath,amssymb}

\usepackage{cite}

\usepackage{nameref,hyperref}
\usepackage{color}

\usepackage{microtype}
\DisableLigatures[f]{encoding = *, family = * }

\usepackage{rotating}

\raggedright
\usepackage[aboveskip=1pt,labelfont=bf,labelsep=period,justification=raggedright,singlelinecheck=off]{caption}

\makeatletter
\renewcommand{\@biblabel}[1]{\quad#1.}
\makeatother

\date{}

\usepackage{lastpage,fancyhdr,graphicx}
\usepackage{epstopdf}
\usepackage{ulem}
\fancyheadoffset[L]{2.25in}
\fancyfootoffset[L]{2.25in}
\definecolor{MyColorBar1}{rgb}{0.9047,0.1918,0.1988}
\definecolor{MyColorBar2}{rgb}{0.2941,0.5447,0.7494}
\definecolor{MyColorBar3}{rgb}{0.3718,0.7176,0.3612}
\definecolor{MyColorBar4}{rgb}{1.0000,0.5482,0.1000}
\definecolor{MyColorBar5}{rgb}{0.8650,0.8110,0.4330}
\definecolor{MyColorBar6}{rgb}{0.6859,0.4035,0.2412}
\definecolor{MyColorBar7}{rgb}{0.9718,0.5553,0.7741}
\definecolor{MySecondColorBar1}{rgb}{0.3467,0.5360,0.6907}
\definecolor{MySecondColorBar2}{rgb}{0.9153,0.2816,0.2878}
\definecolor{MySecondColorBar3}{rgb}{0.4416,0.7490,0.4322}
\definecolor{MySecondColorBar4}{rgb}{1.0000,0.5984,0.2000}
\definecolor{MySecondColorBar5}{rgb}{0.6769,0.4447,0.7114}
\begin{document}
\vspace*{0.35in}

\begin{flushleft}
{\Large
\textbf\newline{Temporal and spatial correlation patterns of air pollutants in Chinese cities}
}
\newline
\\

Yue-Hua Dai \textsuperscript{1,2},
Wei-Xing Zhou\textsuperscript{1,3,4,*}

\bigskip


\bf{1} School of Business, East China University of Science and Technology, Shanghai 200237, China\\
\bf{2} Department of Finance and Management Science, Carson College of Business, Washington State University, Pullman, WA99163, U.S\\
\bf{3} Department of Mathematics, East China University of Science and Technology, Shanghai 200237, China\\
\bf{4} Research Center for Econophysics, East China University of Science and Technology, Shanghai 200237, China\\

\bigskip

%
%




* wxzhou@ecust.edu.cn (WXZ) [{\textit{PLoS ONE}} {\textbf{12}} (8), e0182724 (2017)]

\end{flushleft}
\section*{Abstract}
As a huge threat to the public health, China's air pollution has attracted extensive attention and continues to grow in tandem with the economy. Although the real-time air quality report can be utilized to update our knowledge on air quality, questions about how pollutants evolve across time and how pollutants are spatially correlated still remain a puzzle. In view of this point, we adopt the PMFG network method to analyze the six pollutants' hourly data in 350 Chinese cities in an attempt to find out how these pollutants are correlated temporally and spatially. In terms of time dimension, the results indicate that, except for O$_3$, the pollutants have a common feature of the strong intraday patterns of which the daily variations are composed of two contraction periods and two expansion periods. Besides, all the time series of the six pollutants possess strong long-term correlations, and this temporal memory effect helps to explain why smoggy days are always followed by one after another. In terms of space dimension, the correlation structure shows that O$_3$ is characterized by the highest spatial connections. The PMFGs reveal the relationship between this spatial correlation and provincial administrative divisions by filtering the hierarchical structure in the correlation matrix and refining the cliques as the tinny spatial clusters. Finally, we check the stability of the correlation structure and conclude that, except for PM$_{10}$ and O$_3$, the other pollutants have an overall stable correlation, and all pollutants have a slight trend to become more divergent in space. These results not only enhance our understanding of the air pollutants' evolutionary process, but also shed lights on the application of complex network methods into geographic issues.



\section*{Introduction}
Since 2012, the Chinese government has invested a huge amount of resources in establishing more than 1500 air pollution monitoring centers to dynamically record and publish the air quality index \cite{Tao-2016-AtmosEnv}. However, it still remains a challenge in effectively quantifying how these pollutants evolve across time and cities \cite{Gillespie-Masey-Heal-Hamilton-Beverland-2017-AtmosEnv}, how the occurrence of environmental pollution is temporally and spatially correlated. Nevertheless, measuring the temporal and spatial correlation patterns of air pollutants has a profound significance in understanding cities' connections as well as pollutants' shifting patterns, thus providing a marvelous channel to analyze geographical, meteorological conditions as well as social-economic spillover effect and most importantly, curbing the air pollution with correct remedy \cite{Tao-2016-AtmosEnv,Zhang-Cao-2015-SR}.

To achieve this end, we resort to complex network methodologies and try to quantify the correlations from two aspects. When it comes to the time side, we resort to the fractal analysis to examine city's self-similarity of the six pollutants. As for the space side,  we try to view these cities as scattered nodes and cities' cross correlations of the pollutants time series as the edges in a graph. We then work on extracting the hierarchy structures and refining small correlated groups (known as cliques) in the constructed graph using planar maximally filtered method. Finally, we test the stability of distance and correlation relationship to consolidate our previous analysis. Both correlations are of particular importance in enhancing our understanding of each pollutant's temporal and spatial patterns.

This is one of the few papers trying to understand pollution's correlated patterns from the complex network perspective. Most papers about air pollution have two focuses: Air pollution's causes \cite{He-Tie-Zhang-2015-Particuology,Xie-Tou-Zhang-2016-JClPro} and effects \cite{Tang-Zhao-Wang-2017-AtmosEnv,Liu-Huang-Ma-2017-EnvirInter}. However, our starting point is different in that it serves to deepen the knowledge of each pollutant's evolutionary patterns. In this regard, \cite{Zhang-Cao-2015-SR} do similar work, they also analyze the pollutants' temporal distribution properties at the city level. Our intraday pattern results are partially consistent with theirs. However, their work is like a basic statistical mechanism analysis, which inspires us to deeply mine the latent information. \cite{Tao-2016-AtmosEnv} study the spatial oscillation patterns of six air pollutants. Their research attaches great importance on meteorological conditions and satellite observations, therefore providing reasonable explanations for some of pollutants' intraday patterns and long-term correlations. However, their work differs from ours in several aspects. First, they only analyze air pollution in eastern cities of China in winter, while our analysis covers all the cities of China and four seasons. Second, they successfully explain the air pollution from both geographical and meteorological conditions, while due to data availability we only research the pollution's correlation from geographical distance. Third, they are trying to unveil the spatial oscillation patterns, while our targets are the temporal and spatial correlation structures. In spite of these differences, some of their spatial oscillation patterns are well-identified in our research. Regional-scale temporospatial correlations of air pollutants in China have also been investigated intensively in recent years \cite{Bao-Yang-Zhao-Wang-Yu-Li-2015-IJERPH,Huang-Long-Wang-Huang-Ma-2015-APR,Wang-Ying-Wu-Zhang-Ma-Xiao-2015-RM,Xia-Qi-Liang-Zhang-Jiang-Ye-Liu-Huang-2016-IJGI}. It is found that the spatial and temporal correlation structure is based on regional variations or part of pollutants' variations \cite{Xia-Qi-Liang-Zhang-Jiang-Ye-Liu-Huang-2016-IJGI,Bao-Yang-Zhao-Wang-Yu-Li-2015-IJERPH}. To some extent, the methods used among these papers are quite conventional. Compared to these papers, we transform spatial correlation into spatial network to quantitatively measure the spatial agglomeration or separation. As noted before, our research is the first time to cover almost every medium-sized Chinese cities, making the findings more general and complete.

This paper, on the other hand, is also one of the few trying to apply complex network methods into spatial correlation issues. Although these methods have achieved great success in many areas such as stock market clustering \cite{Tumminello-Aste-DiMatteo-Mantegna-2005-PNAS} and gene decoding \cite{Song-Zhang-2015-PLoS1}, they still remain relatively new in pollution's geographic issues. The physical distances directly or indirectly affect the spatially embedded intercity correlations, making network's architecture radically different from that of random networks \cite{Gastner-Newman-2006-EPJB,Expert-Evans-Blondel-Lambiotte-2011-PNAS}. In this sense, this paper casts a new light on the application of network methods into the pollution's spatial correlation issues.


\section*{Materials and Methods}
\subsection*{Data sets}

We obtained the hourly pollutant data from Shanghai Qingyue Open Environmental Protection Data Center (QOEPDC), and the hourly data contain about 400 observed mainland cities and from 2015-01-01 to 2015-12-31. The six pollutants are PM$_{2.5}$, PM$_{10}$, CO, O$_3$, NO$_2$ and SO$_2$. 

For pollutant $k$ ($k=1,\cdots,6$ are PM$_{2.5}$, PM$_{10}$, CO, O$_3$, NO$_2$ and SO$_2$ respectively), city $i~(i=1,\cdots,350)$'s observed hourly time series $x^{(k)}_{i,t}$ spans the entirety of 2015. Ideally, each time series should have an identical length $T=365\times 24=8760$ hrs. However, due to monitor station recording errors, the actual time series have different lengths less than 8760. In order to preserve data completeness and improve analysis accuracy, we select 350 cities which have more than 8000 observations. Before proceeding to the data analysis, we first check the data quality and find that zeros comprise less than 1\% of the time series. These zeros obviously result from the recording errors and we replace them with the average of corresponding previous and next hour's concentrations. Same alterations are also done to a few extremely high and impossible values.

\subsection*{The temporal correlation structure}

Hurst exponent is a critical variable to quantify whether the trends of air pollutants revert to the mean (low long-term correlations) or to the cluster (high long-range correlations) \cite{Mandelbrot-Wallis-1968-WRR,Mandelbrot-Wallis-1969a-WRR,Mandelbrot-Wallis-1969b-WRR,Mandelbrot-Wallis-1969c-WRR,Mandelbrot-Wallis-1969d-WRR,Kleinow-2002}.
It's defined in terms of the asymptotic behavior of the rescaled range as a function of the time span of a time series,
\begin{equation}
E\left[{R(n)}/{S(n)}\right]=Cn^H~~~~as~~n\rightarrow \infty,
\label{Eq:Hurst:Def}
\end{equation}
where $R(n)$ is the range of the first $n$ values, $S(n)$ is the deviation, $E[\cdot] $ is the expected value and $C$ is a constant. Theoretically, the Hurst exponent $H$ lies between 0 and 1, and is cut off by 0.5: When $0<H<0.5$, the time series is switching between high and low values alternatively (mean-reverting process); while when $0.5<H<1$, it is a long-term dependent process featuring the trend that high value is followed by another higher value. Quantifying the long-range correlation has multiple implications. First, Hurst exponent is an indicator of autocorrelation, which enables us to explore the fractal structure of the evolutionary process. Second, from the policy-making perspective, when one pollution event occurs, more serious pollution events are more likely to happen afterwards, and policy makers thus have a hint to restrict outdoor activities and accordingly fight against the pollution. Last, our estimation of pollutant's Hurst exponent not only enhances our knowledge about air condition's autocorrelation structure but also draws a complete picture about how the strength of autocorrelation of each pollutant in each place differs from the others.

Considering the influence of intraday patterns and possible seasonal variations, we prepare three data sets: The raw data, the normalized data by dividing the hourly average
\begin{equation}
  r^{(k)}_i(d,h)=\frac{x^{(k)}_i(d,h)}{\frac{1}{365}\sum_d x^{(k)}_i(d,h)},
\label{Eq:IP:Detrend:Def}
\end{equation}
and the normalized data by dividing the each season $S$'s hourly average
\begin{equation}
  r^{(k)}_i(d,h)=\frac{x^{(k)}_i(d,h)}{\frac{1}{N_S}\sum_{d\in~{\mathrm{Season}}~S} x^{(k)}_i(d,h)},
\label{Eq:IP:Detrend:SA:Def}
\end{equation}
where $x^{(k)}_i(d,h)$ is pollutant $k$'s concentration level in city $i$ at $h$ on day $d$, $N_s$ is the number of days in season $S~(S=1,2,3,4)$. Because $r^{(k)}_i(d,h)$ is computed on the basis of each city, it automatically eliminates our concerns of the trend issues. In the following part, detrended moving average (DMA) algorithms \cite{Alessio-Carbone-Castelli-Frappietro-2002-EPJB,Carbone-Castelli-2003-SPIE,Carbone-Castelli-Stanley-2004-PA,Carbone-Castelli-Stanley-2004-PRE,Arianos-Carbone-2007-PA,Gu-Zhou-2010-PRE,Jiang-Zhou-2011-PRE,Shao-Gu-Jiang-Zhou-Sornette-2012-SR,Shao-Gu-Jiang-Zhou-2015-Fractals} is applied to compute $H$. The basic idea for DMA algorithms is to remove the trend by considering the second order difference between original time series and its moving average function (detailed procedures can be seen in Refs.~\cite{Vandewalle-Ausloos-1998-EPJB,Gu-Zhou-2010-PRE,Jiang-Zhou-2011-PRE}).

To consolidate how the Hurst exponents are spatially heterogenous, we follow the scheme in Ref.~\cite{Wang-Zhang-Fu-2016-EcolInd} to calculate the spatial stratified heterogeneity. All cities' Hurst exponents $H_{i}$s are stratified into $h=1,2,\cdots,L$ stratum based on socioeconomic factor or geographical factor $H_{i,h}$, and the spatial stratified heterogeneity is defined as
\begin{equation}
q=1-\frac{\sum_{h=1}^L \sum_{i=1}^{N_h} (H_{i,h}-\bar{H}_{h})^2}{\sum_{i=1}^{N} (H_{i}-\bar{H})^2}
\label{Eq:SSH:Def}
\end{equation}
where $N_h$ is the number of cities in stratum $h$ and $\bar{H}_{h}$ is the average Hurst exponent of stratum $h$. As proven in Ref.~\cite{Wang-Zhang-Fu-2016-EcolInd}, a test statistic is constructed as $q$ follows a non-centered $F$ distribution. In this paper, we clarify two kinds of spatial stratified heterogeneities. The first one is stratified by socioeconomic status, where the studied cities are partitioned into 6 groups based on their social economic ability (more information can be retrieved at http://www.stats.gov.cn/english/). The second one is based on geographical locations using the 31 administrative partitions.

\subsection*{The spatial correlation structure}
The Pearson cross correlation is used to quantify the similarity between city $i$ and city $j$ for pollutant $k$ and is defined as
\begin{equation}
\label{Eq:AQI:Correlation:Def}
c^{(k)}_{i,j}=\frac{\sum_{t=1}^T (x^{(k)}_{i,t}-\bar{x}^{(k)}_{i})(x^{(k)}_{j,t}-\bar{x}^{(k)}_{j})}{ \sqrt{\sum_{t=1}^T (x^{(k)}_{i,t}-\bar{x}^{(k)}_{i}})^2 \sqrt{\sum_{t=1}^T (x^{(k)}_{j,t}-\bar{x}^{(k)}_{j})^2}}
\end{equation}
where $\bar{x}^{(k)}_{i}$ is the average concentration of pollutant $k$ in city $i$.

The air quality correlation matrix typically serves as a connection form between these investigated cities which can be viewed as a complex system with interactions and entangles. The correlation matrix has provided crucial information about the system structure. In recent years, the network method and graph theory that incorporate the correlation matrix have increasingly been used to study the complex system from the perspective that observed individual is as the node and the correlation is as the edge linking these individuals \cite{Strogatz-2001-Nature,Albert-Barabasi-2002-RMP,Tumminello-Aste-DiMatteo-Mantegna-2005-PNAS}. The correlation based upon clustering procedure allows us to dig into the hierarchical structure of the system \cite{Tumminello-Aste-DiMatteo-Mantegna-2005-PNAS,Tumminello-Coronnello-Lillo-Micciche-Mantegna-2007-IJBC,Tumminello-Lillo-Mantegna-2010-JEBO,Tumminello-Lillo-Piilo-Mantegna-2012-NJP}. Generally, clustering practically will reduce the dimensions of the researched multivariate time series, and enable us to group the individuals according to the similarity. In this section, we will scrutinize the spatial patterns and cities' intra-cluster structure of each pollutant using the network clustering algorithms based on the correlation matrix.

Tumminello et al. proposed the correlation filtering algorithms by maximizing the planar structure and named it planar maximally filtered graph (PMFG, hereafter) \cite{Tumminello-Aste-DiMatteo-Mantegna-2005-PNAS}. Tumminello, Lillo and Mantegna have compared the several clustering procedures and concluded the PMFG as an extension of minimum spanning tree (MST) that allows loops and cliques in the graph  to provide richer information about the correlation structure \cite{Tumminello-Lillo-Mantegna-2010-JEBO}. The construction procedure for correlation based upon PMFG is rather direct: Starting from the descending sorted list of pair wise correlations $c_{i,j}$, then adding each link between the two cities $i$ and $j$ if and only if the resulting graph can still be embedded on the surface of genus $g\leq k$ after such insertion. The generated simple, undirected, connected graph will have the same hierarchical structure of the minimum spanning tree, but admit loops to retain more relevant information.

\begin{figure}
  \includegraphics[width=0.95\linewidth]{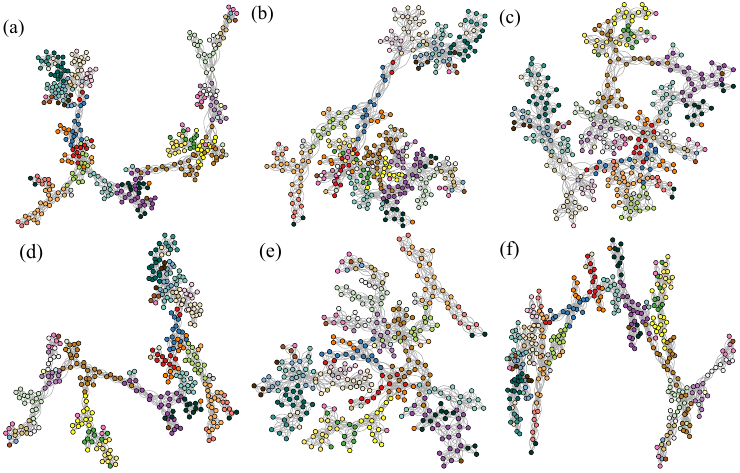}
  \caption{{\bf Large network layout of correlation based upon PMFG. From ({\bf{a}}) to ({\bf{f}}) are PM$_{2.5}$, PM$_{10}$, CO, NO$_2$, O$_3$ and SO$_2$ respectively. To construct the network, we use the $\sqrt{(2-2c_{i,j})}$ as a transformed notation for the correlation.}}
\label{Fig1}
\end{figure}

Fig.~\ref{Fig1} plots the large graph layout for the six pollutants' PMFGs and colors the cities (nodes) in the same province with same color. Same-colored cities tend to be close in geographic distance, and this spatial affinity property is pervasive in the six pollutants. The PMFGs refine many small loops and connected structures known as cliques \cite{Tumminello-Aste-DiMatteo-Mantegna-2005-PNAS}. These cliques are viewed as  small clusters of cities that share high correlations in pollutant's evolutionary dynamics. In the following part, we work on identifying and analyzing these cliques that embedded in the PMFGs.

We also conduct a moving window scheme to analyze how these cliques evolve across the whole year motivated by the fact that the air pollution across the whole country features in dynamics and rebalance resulting from a bunch of geographic, climatic as well as human behavior factors. This study acts as a robustness test over the time stability of refined cliques. To reduce the uncertainty of data length and mitigate the influence of  outliers on our final results, we set the window length $w=720$hr and move forward as $m=24$hr in each step. Within each window, we compute the Pearson correlation matrix as
\begin{equation}
c^{(k)}_{i,j}(t)=\frac{1}{\delta^{(k)}_i \delta^{(k)}_j} \sum_{l=t-w+1}^{t}  [x^{(k)}_{i,l}-\langle{x^{(k)}_{i}}\rangle][x^{(k)}_{j,l}-\langle{x^{(k)}_{j}}\rangle]
\label{Eq:MovingWindows:CorrelationDefinition}
\end{equation}
where $\delta^{(k)}_i$ is the standard deviation of city $i$'s time series of pollutant $k$. In each window we obtain the corresponding correlation-based PMFG networks.

%
%

\section*{Results}
\subsection*{Intraday patterns}
\label{SS:IP}

Understanding the time trend of each pollutant will give us a general view of each pollutant's evolutionary process and help us effectively detrend the time series so as to draw the real correlations\cite{Chen-Ivanov-Hu-Stanley-2002-PRE}. It's natural to start from the intraday patterns due to the fact that the pollutants may be significantly influenced by diurnally cyclical temperature and illumination changes \cite{Mayer-1999-AtmosEnv,Panday-Prinn-2009-JGR}. These intraday patterns showed in a daily periodical phenomenon have been found pervasive in natural sciences such as temperature variability \cite{Pardo-Meneu-Valor-2002-EE,Keggenhoff-Elizbarashvili-King-2015-WCE}, rainfall perception \cite{Buytaert-Celeri-Willems-DeBievre-Wyseure-2006-JH} and social sciences such as market trading activities \cite{Admati-Pfleiderer-1988-RFS,Mcinish-Wood-1992-JF,Gu-Chen-Zhou-2007-EPJB}, human mobilities \cite{Jiang-2012-DMKD}. In this section, we begin with parsimonious models to display the intraday patterns of the six pollutants.

Let $x^{(k)}(d,h)$ denote the 350 cities' averaged concentration of pollutant $k$ at the $h$-hour on day $d$. The normal definition of intraday patterns is as follows:
\begin{equation}
  x^{(k)}(h) = \frac{1}{365}\sum_{d=1}^{365} x^{(k)}(d,h),
  \label{Eq:IP:xi:normal}
\end{equation}
which averages the pollutant concentrations at the same hours of all the days.

An alternative definition reads:
\begin{equation}
  x^{(k)}(h) = \frac{1}{365}\sum_{d=1}^{365} \frac{x^{(k)}(d,h)}{\max\limits_h\{x^{(k)}(d,h)\}},
  \label{Eq:IP:xi:max}
\end{equation}
where $\max\limits_h\{x^{(k)}(d,h)\}$ is the maximum value of pollutant $k$ on day $d$. This definition takes into account the seasonal variation of pollutant concentration and rescales the concentration with respect to the maximum value on each day.

A third definition reads:
\begin{equation}
  x^{(k)}(h) = \frac{1}{365}\sum_{d=1}^{365} \frac{x^{(k)}(d,h)}{\frac{1}{24}\sum_{h=1}^{24} x^{(k)}(d,h)},
  \label{Eq:IP:xi:mean}
\end{equation}
where $\frac{1}{24}\sum_{h=1}^{24} x^{(k)}(d,h)$ is the average value of pollutant $k$ on day $d$. This definition also takes into account the seasonal variation of pollutant concentration, but rescales the concentration with respect to the average value on each day.

These three definitions commonly present the intraday patterns but differ in relative magnitude. Eq.~(\ref{Eq:IP:xi:normal}) retains the original unit and magnitude, while Eq.~(\ref{Eq:IP:xi:max}) and Eq.~(\ref{Eq:IP:xi:mean}) scale the raw data by dividing that day's maximal or mean concentration.

\begin{figure}[htb]
  \centering
  \includegraphics[width=0.95\linewidth]{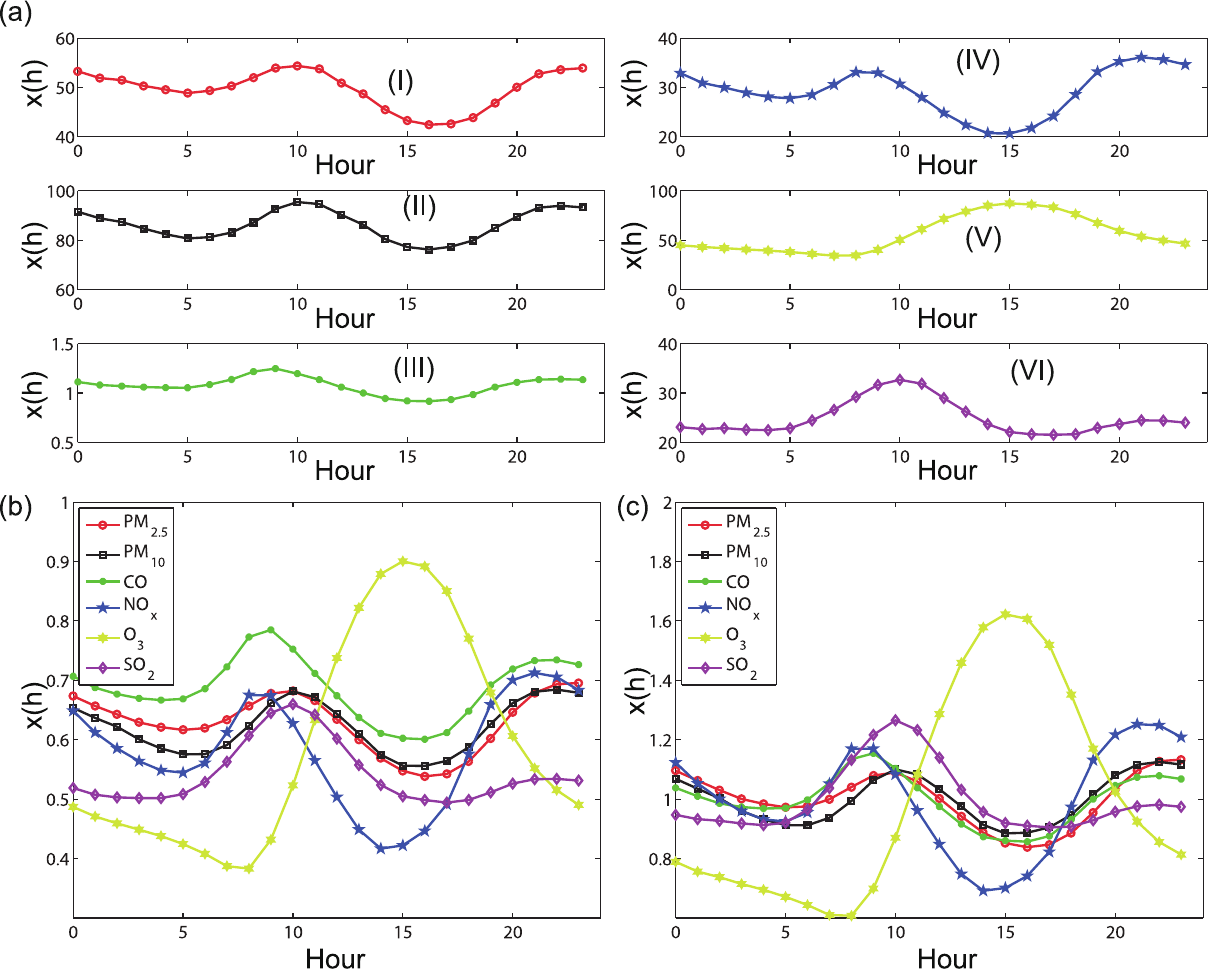}
  \caption{{\bf Intraday patterns of six pollutants measured by three definitions. The results in plots ({\bf{a}}), ({\bf{b}}) and ({\bf{c}}) correspond respectively Eq.~(\ref{Eq:IP:xi:normal}), Eq.~(\ref{Eq:IP:xi:max}) and Eq.~(\ref{Eq:IP:xi:mean}). Each hour is averaged using 350 cities' sample across the 365 days in 2015. In ({\bf{a}}), except CO's unit is $mg/m^3$, other five pollutants' units are all $\mu g/m^3$, and from {\bf{I}} to {\bf{VI}} are PM$_{2.5}$, PM$_{10}$, CO, NO$_2$, O$_3$ and SO$_2$ respectively. }}
\label{Fig2}
\end{figure}

Fig.~\ref{Fig2} shows the three defined intraday patterns by averaging the 350 observed cities' concentrations in each hour. Generally, each pollutant's averaged concentration time series is featured in cyclical patterns within one day. Except for O$_3$, the other five pollutants' intraday patterns are composed of two contraction periods (from 12 AM to 5 AM and from 10 AM to 15 PM) and two expansion periods (from 5 AM to 10 AM and from 15 PM to 23 PM). These five pollutants' concentrations simultaneously hike to the peak level around 10 AM and then reduce to the lowest level around 15 PM. These fluctuations imply the ``periodic'' daily human activities because NO$_2$ and SO$_2$ mainly come from vehicles and coal combustion. However, O$_3$'s concentration continues reducing until 9 AM and then bounds to the peak level around 15 PM due to the photochemical reaction\cite{Kassem-2014-WE}. The peak time for O$_3$ is a trough time for the other five pollutants. After 15 PM O$_3$ is on the way to decline until midnight. The three definitions share almost identical intraday patterns and differ in relative magnitudes. Another conspicuous discrepancy lies in the relative volatility within one day. In Fig.~\ref{Fig2}(c), the concentration of O$_3$ has the highest volatility, and NO$_2$ ranks the second while other four pollutants have only $\pm 0.1$ relative change around the mean level.

\begin{figure}[htb]
  \includegraphics[width=0.95\linewidth]{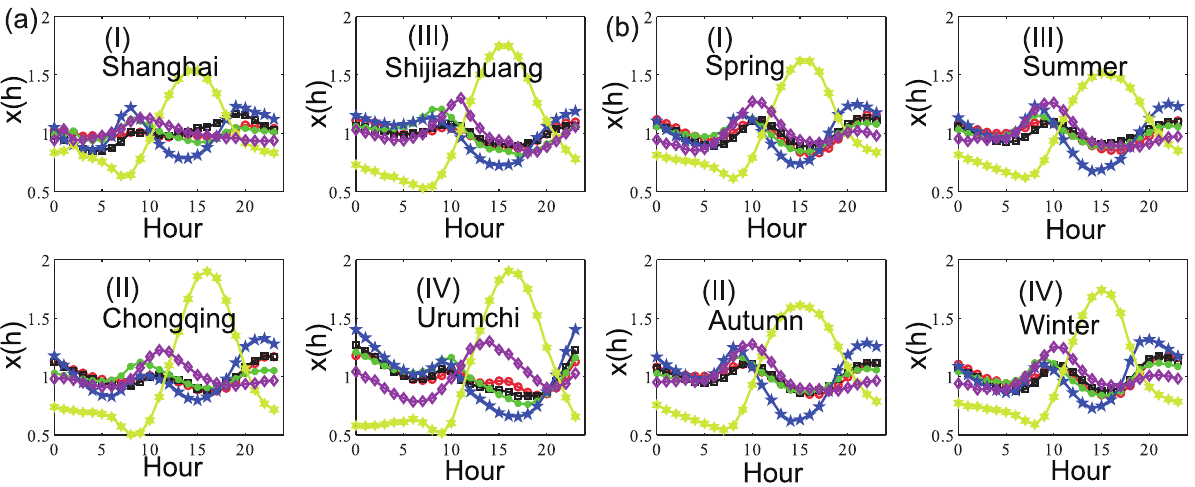}
  \caption{{\bf Intraday patterns defined in Eq.~(\ref{Eq:IP:xi:mean}) of air pollutants in four selected cities ({\bf{a}}) and four seasons ({\bf{b}}). Each city's intraday patterns ({\bf{a}}) are averaged using the sample across the 365 days in 2015. The four cities are Shanghai ({\bf{I}}), Chongqing ({\bf{II}}), Shijiazhuang ({\bf{III}}) and Urumchi ({\bf{IV}}); Each season's intraday patterns ({\bf{b}}) are averaged using the 350 cities in the three season months (spring ({\bf{I}}) is from January to March, summer ({\bf{III}}) is from April to June, autumn ({\bf{II}}) is from July to September and winter ({\bf{IV}}) is from October to December). The meaning of the line types refers to Fig.~\ref{Fig2}({\bf{b}}) and ({\bf{c}}).}}
  \label{Fig3}
\end{figure}

There are still two concerns about the robustness of air pollutants' intraday patterns. One is whether each individual city shares the same intraday trends as the aggregated does in Fig.~\ref{Fig2}, the other is whether the aggregated intraday patterns are persistent during the four seasons. Fig.~\ref{Fig3}(a) displays four typical cities' intraday patterns: Shanghai, Chongqiong, Shijiazhuang and Urumchi (the four cities are in different geographic areas and economic zones, also represent the four development levels of Chinese cities) and (b) averages the hourly level within each season. Both figures show a consistent framework with the previous studies. They specifically differ in relative magnitudes. For example, Fig.~\ref{Fig3}(a) shows that the six pollutants in Shanghai generally fluctuate more steadily than other cities do and Shanghai also has a relatively low pollutants level. To a large extent, this is determined by Shanghai's location and its service-oriented economy. Shijiazhuang and Urumchi are both highly polluted cities, but the sources of pollutants in the two cities are quite different. Shijiazhuang's intensive heavy industry is the leading cause and Urumchi's location and climatic causes outweigh others. Fig.~\ref{Fig3}(b) shows the intraday patterns across the four seasons are almost identical. The subtle difference resides in the minimum NO$_2$ level, which is a bit lower in summer and autumn than that of spring and winter. These findings, both at the city level and the season level, are quite consistent to Ref.~\cite{Zhang-Cao-2015-SR}'s summarized results regardless of adopting different data sets and sample cities. These roughly-constructed but well-identified intraday patterns inspire us to scrutinize each pollutant's time series periodicity in a detailed way.

\subsection*{Lomb power analysis}

In this section, we introduce the normalized Lomb power \cite{Lomb-1976-ApSS,Ni-Zhou-2009-JKPS} to confirm the cyclic patterns that have been captured in Fig.~\ref{Fig2} and Fig.~\ref{Fig3}. Similar to Fourier transformation, Lomb power analysis works on converting the cyclic time series into frequency domain so as to obtain the periodical parameters. For evenly sampled time series, Lomb power is equivalent to conventional Fourier transformation spectrum analysis. For unevenly sampled time series, Lomb power analysis performs better by effectively mitigating the long-periodic noise caused by long gapped records \cite{Lomb-1976-ApSS}. The Lomb power $P_T(f)$ is defined as
\begin{equation}
  P^{(k)}_T(f)=\frac{1}{2{\sigma^{(k)}}^2}\left\{\frac{[\sum_s (x^{(k)}_s-\bar{x}^{(k)})\cos2\pi f(t_s-\tau)]^2}{\sum_s \cos^2 2\pi f(t_s-\tau) }+\right.
  \left.\frac{[\sum_s (x^{(k)}_s-\bar{x}^{(k)})\sin 2\pi f(t_s-\tau)]^2}{\sum_s \sin^2 2\pi f(t_s-\tau) }\right\},
\end{equation}
where $x^{(k)}_s(s=1,\cdots,8760)$ is the averaged time series of pollutant $k$ with size $T=8760$, $\bar{x}^{(k)}$ and $\sigma^{(k)}$ are the mean and standard deviation of the time series, and the time offset $\tau$ is determined by
\begin{equation}
\tan (2\pi f\tau)=\frac{\sum_s \sin(2\pi f t_s)}{\sum_s \cos(2\pi f t_s)}.
\end{equation}

\begin{figure}
  \includegraphics[width=0.95\linewidth]{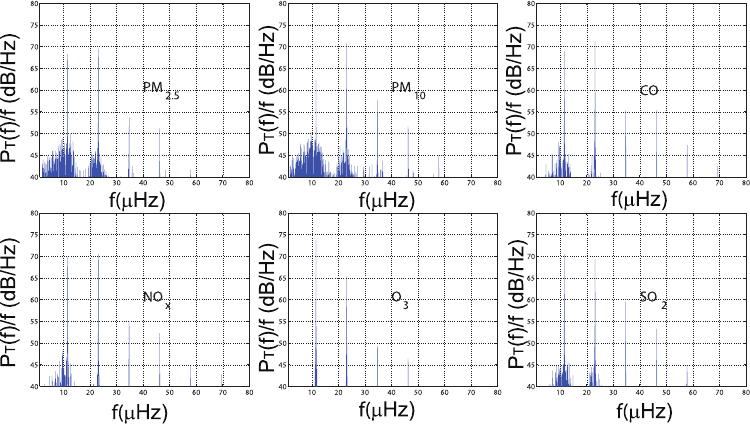}
  \caption{{\bf Normalized Lomb power spectrum $P^{(k)}_T(f)$ for each averaged pollutant concentration time series $x^{(k)}$.}} 
\label{Fig4}
\end{figure}

Fig.~\ref{Fig4} displays the Lomb periodograms of six pollutants time series. Obviously, the six time series share an almost identical peak power around $f=11.58~\mu Hz$ and $P_T(f)=68.31~dB/HZ$, which equals to a period of 23.99 hrs and is corresponding to the diurnal pattern of the pollutants \cite{Ni-Zhou-2009-JKPS}. Except for O$_3$, $2f$ is also a peak level for the Lomb power, and even higher than the first peak, which is explained by the intraday cycles noted before: Within one day, the evolutionary characteristic of the pollutants is viewed as two cycles, and the second peak is corresponding to such an approximately half day period. Another straightforward feature shown in Fig.~\ref{Fig4} is the evenly spaced harmonic peaks, they serve to consolidate the intraday patterns and these patterns can be safely decomposed into two contraction and two expansion periods. Moreover, this decomposition is both statistically significant and intuitively reasonable. 

In this section, we have satisfactorily uncovered each pollutant's intraday patterns: From both city level and season level, two regimes within one day are identified. However, O$_3$ is an exception because of its asynchronous changing time with other pollutants. The Lomb power analysis enhances our understanding of this periodicity from power spectrum peaks. Practically, these intraday patterns recognized as important air quality evolutionary clues may be of great value in scheduling outdoor activities \cite{Wang-Ying-Wu-2015-RM} and controlling air pollution \cite{Zhang-Cao-2015-SR}.

\subsection*{Temporal correlation structure}

\begin{figure}
  \includegraphics[width=0.95\linewidth]{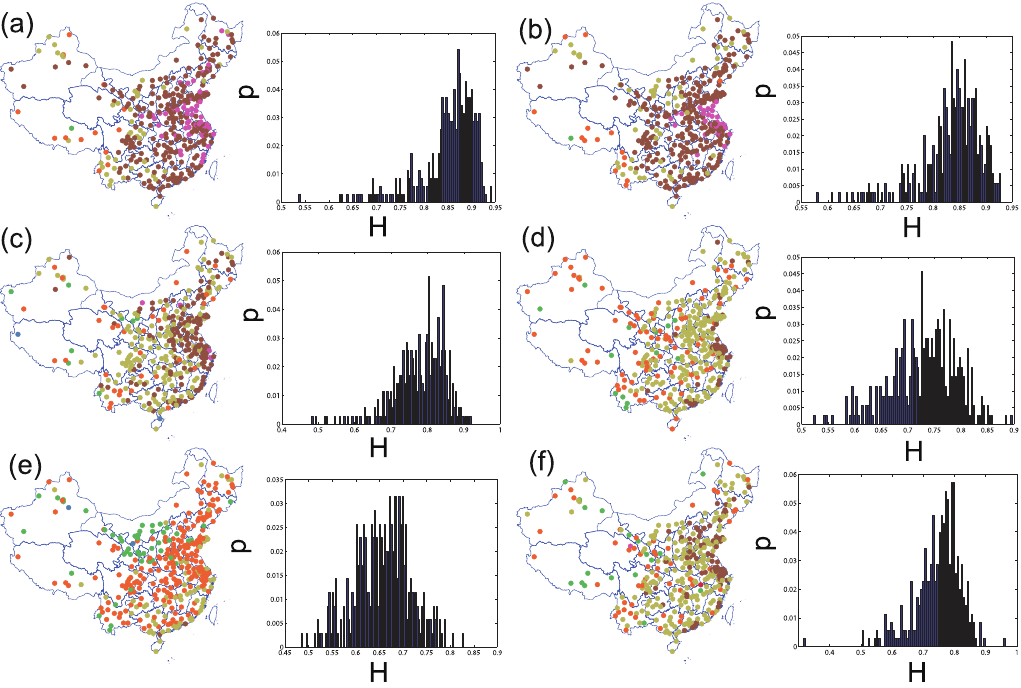}
  \caption{{\bf Hurst exponent distribution of the six pollutants for the 350 observed cities in 2015. From ({\bf{a}}) to ({\bf{f}}) are PM$_{2.5}$, PM$_{10}$, CO, NO$_2$, O$_3$ and SO$_2$ respectively. The left panel for each subfigure are the Hurst exponents projected into the China maps, in which the the filled circles in different colors (red, blue, green, orange, yellow, brown and pink) are 7 equal Hurst intervals ranging from 0.3 to 1. The right panel for each subfigure are the histograms of the exponents.}}
  \label{Fig5}
\end{figure}

We adopt the DMA method to estimate the Hurst exponent of each pollutant time series in each place. Then, we project these Hurst exponents into the Chinese map and also plot the histograms to illustrate their distributions.
Fig.~\ref{Fig5} shows the raw data's Hurst exponents distributions both in geographical style (left panel) and histogram style (right panel). Generally, most of researched time series have Hurst exponents significantly higher than 0.50, which signals strong long-term correlations for air pollution. This provides solid evidence for the phenomenon that smoggy days are always followed by one after another. Geographically neighboring cities tend to share similar long-term correlations in Fig.~\ref{Fig5}'s left panel of each plot. The six pollutants' exponents show the feature of unimodal and asymmetric distributions in the right panel. Specifically, Fig.~\ref{Fig5}(a) and (b) apparently display that PM$_{2.5}$ and PM$_{10}$ enjoy the highest average $H$s ( $\bar{H}_{PM_{2.5}}=0.85$ and $\bar{ H}_{PM_{10}}=0.83$) and the distributions are left-skewed with the mode around 0.87. The highest long-term correlated areas span from Bohai Bay to Fujian Province vertically and from Heinan Province to Shanghai horizontally. The average exponents of  CO, NO$_2$, O$_3$, SO$_2$ are 0.78, 0.73, 0.66, 0.75 respectively and have more symmetric distributions. The most strongly long-term correlated areas for the four pollutants locate in the east coastal part and center on the Yangtze River Delta. However, most cities' $H$s of O$_3$ lie between 0.50 and 0.60, much lower than other pollutants, especially for the north part of China. It's rather difficult to distinguish most cities' O$_3$'s concentration time series from random walk process. In other words, it's not easy to track O$_3$'s long-term dynamic patterns. Finally,  although the estimated exponents are sensitive to parameters used in DMA algorithms and even sensitive to the algorithms used \cite{Xie-Jiang-Zhou-2014-EM}, there are still five cities' $H$s are appreciably lower than 0.50: Haikou (in Hainan Province, $110.32^{\circ}$E, $20.03^{\circ}$N) and Ali Area (in Tibet Province,$80.10^{\circ}$E, $32.50^{\circ}$N)'s CO series; Zhongwei (in Ningxia Province, $105.18^{\circ}$E, $37.52^{\circ}$N) and Turpan Area (in Xinjiang Province, $89.17^{\circ}$E, $42.95^{\circ}$N)'s O$_3$ series; Jiujiang (in Jiangxi Province, $116.00^{\circ}$E, $29.70^{\circ}$N)'s SO$_2$ series.

The high long-term correlations of the pollutants (except for O$_3$) are partly consistent with the successive random dilution (SRD) explanation \cite{Ott-1994,Lee-2002-WASP}. An initially concentrated pollutant will experience a random dilution and mixing process in the air, of which the process is lognormally distributed. As stressed in Ref.~\cite{Lee-2002-WASP}, the extreme concentration variability in time, with intensity peaks many times higher than the average, may be viewed as a consequence of a multiplicative dilution process. On the other hand, the air pollution time series possess high long-term correlations like other natural phenomenon such as river overflow and rainfall perception, since the periodic impact originated from human activity is bound to contribute some long-term correlations. In view of this, it's quite important to restrict the emissions once a pollution event happens, and sufficient time for the random dilution process will change the long-term structure of air pollutants.

As observed from Fig.~\ref{Fig5}, two fine particulate matters share similar spatial and probabilistic distribution properties of $H$s. To find out how the pollutants' $H$s are correlated, we tabulate the correlation coefficients of any two pollutants' Hurst exponents in Table~\ref{Tab1}. The table shows that one pollutant's long-term correlation properties have a positive link with that of another pollutant, especially for the two fine particulate matters. This high correlation between pollutants stems from two possibilities: First, all the observed cities are commonly influenced by the wave of pollution with another wave of pollution following after, which results in positive correlations between the pollutants. Secondly, even if the pollution doesn't occur simultaneously, the regional and asynchronous pollution periods are linked with each other through some methods or driven by some common causes. For example, PM$_{2.5}$ and PM$_{10}$ have the highest correlation of all the pairs due to the similar source of the two pollutants. O$_3$'s Hurst exponents generally correlates weakly with other pollutants, which is consistent to the previous finding that the peak time of O$_3$ is the trough time of other pollutants and O$_3$'s distribution is quite different from others. In other words, this asynchronism reduces correlation between the O$_3$'s Hurst exponents and that of other pollutants. However, there are still some overlapped contraction and expansion periods between O$_3$ and other pollutants, which ensures the correlations are still positive. Except for above-mentioned pollutants, the other three pollutants' Hurst exponents maintain the correlation level ranging from 0.4 to 0.5, a moderate strength of positive correlation.

Inspired by \cite{Wang-Haining-Liu-Li-Jiang-2013-EPA,Wang-Zhang-Fu-2016-EcolInd}, we measure the heterogeneity stratified by the social economic indicator (labeling each city 1-6 based on its socioeconomic development level) and the geographic indicator (using administrative partition as the indicator). The results show that the Hurst exponents are more spatially homogenous in terms of the socioeconomic partitions than geographic partitions. In other words, cities belong to the same development level tend to share similar long-term correlation structure in pollutant time series. This conclusion has several implications. First, cities within the same development level are more likely to share similar energy-intensive heavy industry structure \cite{Luo-Chen-Zhu-Peng-Yang-Yang-Zhang-2014-PLoS1}. Second, the geographical closeness means a similar air pollution mixing and diluting ability. Therefore, to curb the emission, this heterogeneity will inspire us to choose a cooperative model in an effective way.

\begin{table}[htbp]
\centering
\caption{{\bf Correlations and spatial stratified heterogeneity of the six pollutants' Hurst exponents}}
\setlength\tabcolsep{4pt}
\scriptsize
\begin{tabular}{ccccccc}
\hline
Pollutants &PM$_{2.5}$&PM$_{10}$&CO&NO$_2$&O$_3$&SO$_2$\\
\hline
\multicolumn{7}{c}{Panel A: Correlations of Hurst exponents}\\
\hline
PM$_{2.5}$& 1.00  & 0.88  & 0.53  & 0.57  & 0.19  & 0.52  \\
PM$_{10}$&  0.88  & 1.00  & 0.54  & 0.61  & 0.23  & 0.48  \\
CO& 0.53 & 0.54  & 1.00  & 0.60  & 0.29  & 0.52  \\
NO$_2$& 0.57  & 0.61  & 0.60  & 1.00  & 0.41  & 0.55  \\
O$_3$& 0.19  & 0.23  & 0.29  & 0.41  & 1.00  & 0.22  \\
SO$_2$& 0.52  & 0.48  & 0.52  & 0.55  & 0.22  & 1.00  \\
\hline

\multicolumn{7}{c}{Panel B: Spatial stratified heterogeneity of Hurst exponents}\\
\hline
Socioeconomic stratum&0.09  & 0.06  & 0.01  & 0.07  & 0.00  & 0.09 \\
Geographical stratum&0.57  & 0.60  & 0.41  & 0.55  & 0.52  & 0.40\\
\hline
\end{tabular}
\label{Tab1}
\end{table}

\begin{table}[htbp]
\centering
\caption{{\bf Correlations between Hurst exponents and mean, standard deviation, skewness and kurtosis across 350 cities of each pollutant. This table reports the raw pollutants data results (Panel A), the intraday detrended data results (Panel B) and seasonal adjusted detrended data results (Panel C).}}
\setlength\tabcolsep{4pt}
\scriptsize
\begin{tabular}{ccccccc}
\hline
 &PM$_{2.5}$&PM$_{10}$&CO&NO$_2$&O$_3$&SO$_2$\\
 \hline
 \multicolumn{7}{c}{Panel A: Raw data} \\
 \hline
 $c$(Mean,$H$)&0.41  & 0.36  & -0.12  & 0.46  & 0.20  & 0.07  \\
 $c$(Std,$H$)&0.25  & 0.15  & -0.21  & 0.47  & -0.14  & -0.14  \\
 $c$(Skewness,$H$)&-0.23  & -0.20  & -0.05  & -0.15  & 0.03  & -0.28  \\
 $c$(Kurtosis,$H$)& -0.20  & -0.20  & -0.08  & -0.07  & 0.09  & -0.21  \\
 \hline
 \multicolumn{7}{c}{Panel B: Intraday detrended data} \\
 \hline
$c$(Mean,$H$)&0.03  & -0.07  & 0.01  & 0.07  & -0.06  & -0.04  \\
$c$(Std,$H$)&-0.04  & -0.16  & -0.22  & -0.01  & 0.02  & -0.18  \\
$c$(Skewness,$H$)&-0.18  & -0.24  & -0.03  & -0.11  & 0.08  & -0.30  \\
$c$(Kurtosis,$H$)&-0.15  & -0.23  & -0.09  & -0.11  & -0.04  & -0.23  \\
    \hline
\multicolumn{7}{c}{Panel C: Intraday detrended and seasonal adjusted data} \\
    \hline
$c$(Mean,$H$)&-0.02  & -0.10  & -0.26  & -0.05  & 0.09  & -0.33  \\
$c$(Std,$H$)& -0.14  & -0.18  & -0.42  & -0.19  & -0.13  & -0.48  \\
$c$(Skewness,$H$)&-0.27  & -0.31  & -0.09  & -0.15  & 0.05  & -0.42  \\
$c$(Kurtosis,$H$)&-0.20  & -0.23  & -0.09  & -0.06  & 0.13  & -0.32  \\
\hline
\end{tabular}
\label{Tab2}
\end{table}

As noted before, Hurst exponent is a critical statistic to measure the long-term memories of the air pollutants. As other commonly used basic statistics, Hurst exponent could reflect the trend of the time series. Many papers have documented the potential relationship between Hurst exponent and some summary statistics \cite{Mitra-2012-AsianSS}. Here we assess the connections between Hurst exponents and four basic statistics (mean, standard deviation, skewness and kurtosis) so as to consolidate air pollutants' temporal correlations.

Table~\ref{Tab2} reports mixed results about the relationship between Hurst exponents and the four basic statistics. The Hurst exponent is strongly correlated with the first and second moments of raw data, but if the pollution concentration's daily pattern is detrended (in Panel B and C), the correlation reduces to a very low level (even anti-correlation). Another interesting finding is the correlation between Hurst exponents and skewness or kurtosis. Skewness is a measure of the asymmetry of distribution and kurtosis measures the tailedness of the distribution. Except O$_3$, the other pollutants show a negative relationship between Hurst exponents and the two statistics, in terms of the skewness. It can be interpreted as for more negative skewed pollutants distributions (more high pollutant levels than low levels), the time series tend to be higher in $H$ due to higher likelihood of one polluted day is followed by another more polluted day. However, O$_3$ is an exception due to the commonly low levels of Hurst exponents and its disordered spatial distribution. As for the kurtosis, it's similar to the skewness in that most of the variance results from infrequent extreme deviations, thus leading to a lower Hurst exponent in pollutant level.

So far, we have concluded the pollutants' temporal characteristic as long-term correlation with variations existing between cities and pollutants. This long-term correlation trend is a reasonable explanation for the fact that the polluted days always recurred in clusters. By linking with time series' basic statistics, we find except for O$_3$, the other five pollutants' skewness and kurtosis are negatively related with their Hurst exponents.

\subsection*{Spatial correlation structure and refined cliques}

\begin{figure}
  \includegraphics[width=0.95\linewidth]{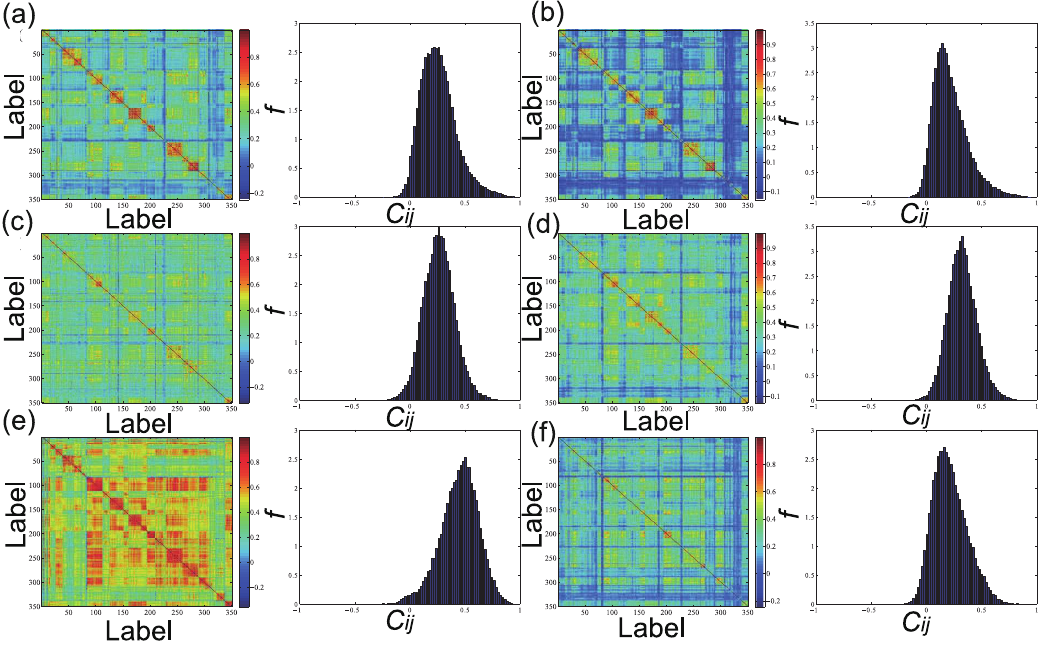}
  \caption{{\bf Correlation spectrum (left panel) and probability density (right panel) of the six pollutants in 350 observed cities. In each subfigure, from (a) to (f) are PM$_{2.5}$, PM$_{10}$, CO, NO$_2$, O$_3$ and SO$_2$ respectively.}}
\label{Fig6}
\end{figure}

Fig.~\ref{Fig6} presents the spectrum and probability distribution of the correlation coefficients of raw data. Obviously the correlation spectrum are featured by clusters. As we arrange our city label sequence according to provincial administrative divisions, cities in the correlation spectrum agglomerate based on their short geographic distance or identical administrative division. These clusters reflect a similar evolutionary process between cities and this similarity is a integrated result of natural conditions and human activities. Generally, neighbouring cities are more likely to share similar meteorological and terrain conditions, as well as development levels.  Therefore, the spectrum shows very straightforward squared clustering patterns. Another point for the overall correlation is they are all appreciably positive which shows the general co-movement of the air pollution across the major cities in China \cite{Chan-Yao-2008-AtmosEnv}. Even so, the six pollutants still possess specific different correlation patterns: The overall average correlation of O$_3$ ($\bar{c}=0.45$ in Table~\ref{Tab3}) is much higher than the other pollutants, which implies that O$_3$'s evolutionary patterns all over Chinese cities is much homogenous. Second to O$_3$, the NO$_2$'s averaged correlation is about $0.32$ and the other three pollutants share nearly similar average correlations from 0.20 to 0.26.

The unimodal distribution for the six correlations presents us the scenario that for the correlation structure, both independent and  perfect correlation are unlikely to happen. The mode correlations varies from 0.25 to 0.50, when checking the lowest correlations, we find these low correlated areas are usually in Yunnan Province and Ningxia Province, of which the cities in these regions enjoy advantageous natural conditions that free them from the outside pollution. In addition, these cities have a relative low percentage of industrialization, preventing them from releasing excessive pollutants. Therefore, they take on an isolated effect from other areas in terms of the air pollution.

\begin{table}[htbp]
\centering
\caption{{\bf Summary statistics of the correlation coefficients of pollutants}}
\setlength\tabcolsep{4pt}
\scriptsize
\begin{tabular}{rrrrrrrr}
 \hline
  Pollutant & Max&Min&Mean&Median&Std&Skewness&Kurtosis\\
   \hline
   \multicolumn{8}{c}{Panel A: Raw data} \\
   \hline
   PM$_{2.5}$& 0.94  & -0.25 & 0.26  & 0.24  & 0.16  & 0.73  & 3.70 \\
   PM$_{10}$&0.93  & -0.16 & 0.22  & 0.19  & 0.15  & 0.95  & 4.08 \\
   CO& 0.84  & -0.33 & 0.26  & 0.26  & 0.14  & 0.14  & 3.25 \\
   NO$_2$& 0.90  & -0.15 & 0.32  & 0.32  & 0.13  & 0.24  & 3.26 \\
    O$_3$&0.95  & -0.36 & 0.45  & 0.46  & 0.17  & -0.41 & 3.24 \\
   SO$_2$& 0.87  & -0.24 & 0.21  & 0.19  & 0.15  & 0.50  & 3.00 \\
   \hline
   \multicolumn{8}{c}{Panel B: Intraday detrended data} \\
   \hline
     PM$_{2.5}$&0.94  & -0.27 & 0.25  & 0.23  & 0.16  & 0.68  & 3.59 \\
    PM$_{10}$& 0.93  & -0.17 & 0.21  & 0.18  & 0.16  & 0.93  & 4.01 \\
    CO&0.84  & -0.35 & 0.25  & 0.25  & 0.15  & 0.12  & 3.24 \\
    NO$_2$&0.89  & -0.23 & 0.28  & 0.28  & 0.14  & 0.15  & 3.19 \\
    O$_3$&0.93  & -0.52 & 0.24  & 0.24  & 0.19  & -0.03 & 2.92 \\
    SO$_2$&0.86  & -0.26 & 0.21  & 0.19  & 0.16  & 0.49  & 2.99 \\
   \hline
   \multicolumn{8}{c}{Panel C: Intraday detrended and seasonal adjusted data} \\
   \hline
    PM$_{2.5}$&0.92  & -0.43 & 0.13  & 0.11  & 0.18  & 0.73  & 3.73 \\
    PM$_{10}$&0.93  & -0.36 & 0.12  & 0.09  & 0.17  & 0.84  & 3.91 \\
    CO&0.83  & -0.59 & 0.11  & 0.11  & 0.19  & 0.07  & 3.19 \\
    NO$_2$&0.86  & -0.42 & 0.20  & 0.19  & 0.17  & 0.14  & 3.08 \\
    O$_3$&0.96  & -0.46 & 0.48  & 0.50  & 0.21  & -0.69 & 3.41 \\
    SO$_2$&0.79  & -0.19 & 0.12  & 0.11  & 0.10  & 0.99  & 4.62 \\
\hline
    \end{tabular}
  \label{Tab3}
  \end{table}

\begin{table}
	\centering
	\caption{{\bf Statistical properties of 3-clique and 4-clique structure. This table reports the summary statistical properties of the 3-clique and 4-clique subgraphs extracted from the correlation matrix based upon Planar Maximally Filtered Graph (PMFG). For each pollutant we report the number of the 3-clique and  4-clique subgraphs belongs to 1-4 administrative provinces respectively.}}
	\label{Tab4}
	\setlength\tabcolsep{4pt}
	\scriptsize
	\begin{tabular}{ccccc}
		\hline
		Pollutant &1 province & 2 provinces & 3 provinces & 4 provinces\\
		\hline
		\multicolumn{5}{c}{Panel A: 3-clique summary }\\
		\hline
		PM$_{2.5}$&  588   & 404   & 50    &  \\
		&(56.43\%) &  (38.77\%) &  (4.80\%)  &  \\
		PM$_{10}$&569   & 394   & 58    &  \\
		&(55.73\%) &  (38.59\%) &  (5.68\%)  &  \\
		CO&408   & 417   & 204   &  \\
		&(39.65\%) &  (40.52\%) &  (19.83\%) &  \\
		O$_3$&530   & 425   & 83    &  \\
		&(51.06\%) &  (40.94\%) &  (8.00\%)  &  \\
		NO$_2$&482   & 453   & 89    &  \\
		&(47.07\%) &  (44.24\%) &  (8.69\%)  &  \\
		SO$_2$& 309   & 505   & 226   &  \\
		&(29.71\%) &  (48.56\%) &  (21.73\%) &  \\
		\hline
		\multicolumn{5}{c}{Panel B: 4-clique summary }\\
		\hline
		PM$_{2.5}$&175&142&29&1\\
		& (50.43\%) & (40.92\%) & (8.36\%)  & (0.29\%) \\
		PM$_{10}$& 160   & 131   & 29    & 2 \\
		& (49.69\%) & (40.68\%) & (9.01\%)  & (0.62\%) \\
		CO& 110   & 123   & 61    & 35 \\
		& (33.43\%) & (37.39\%) & (18.54\%) & (10.64\%) \\
		O$_3$&  150   & 145   & 41    & 6 \\
		& (43.86\%) & (42.40\%) & (11.99\%) & (1.75\%) \\
		NO$_2$& 129   & 142   & 48    & 4 \\
		&(39.94\%) & (43.96\%) & (14.86\%) & (1.24\%) \\
		SO$_2$& 77    & 150   & 85    & 32 \\
		& (22.38\%) & (43.60\%) & (24.71\%) & (9.30\%) \\
		\hline
	\end{tabular}
\end{table}

Table~\ref{Tab4} reports the percentages of 3-clique and 4-clique structures that distribute within 1 identical province or among 2-4 different provinces. It measures the diffusion power of the small similar structures.  In the six pollutants, except SO$_2$, about half of three highly correlated cities are just limited to one province and another 40\% spreads to another province, very few of which (only around 10\%) have the far reaching power to expand to the third province. Cities' SO$_2$ cliques, however, are more likely to stay in adjacent two provinces. The 4-clique has a similar pattern to the 3-clique: The percentage of 4-cliques distributed in 4 distinguished provinces comprises less than 10\%, and most of the 4-cliques stay in the same province or their adjacent province. Cross-sectionally, PM$_{2.5}$ and PM$_{10}$ have the most 1-province cliques and SO$_2$ the least, which is consistent with the previous correlation spectrum that SO$_2$ is the least localized pollutant due to its source of fossil fuel combustion at power plants. The clique distribution, to some extent, resulting from integrated results of local emissions and global transmission. In this sense, the six pollutants can be sorted into 3 groups. First, particulate matters (PM$_{2.5}$ and PM$_{10}$) have the lowest transmission power. The most correlated community for these matters to transfer is within 2 provinces, considering that these matters mainly come from traffic emission and dust \cite{Dan-Zhuang-Li-Tao-Zhuang-2004-AtmosEnv}. It's vital to restrict traffic emission and improve city green land area \cite{Huang-Long-Wang-Huang-Ma-2015-APR}. O$_3$ and NO$_2$ come as the second group in terms of the dispersion power. As noted before, sunlight is tightly associated with the two pollutants \cite{Azmi-Latif-Ismail-Juneng-Jemain-2010-AQAH,Kassem-2014-WE}. Hence, controlling the concentration of these two pollutants should mainly focus on its heavy industry emission locally. As for the other two pollutants, regional control is far from enough, interregional cooperation would be more effective than the local's effort.

Table~\ref{Tab5} tabulates the strongest correlated 3-clique and 4-clique of each pollutant. In panel A, the highest correlated 3-cliques of PM$_{10}$, NO$_2$ and O$_3$  are in the same province, and the three cities in 3-clique of  PM$_{2.5}$ and CO actually belong to the city group in Yangtze River Delta. Interestingly, the highest correlated 3-clique for SO$_2$ is composed of the three capital cities in Northeast China, which are viewed as historical industrial bases, as evinced that SO$_2$ is not a local pollutant and its variations are highly connected because of industrial activities. The extended 4-clique results of Table~\ref{Tab5} panel B differ from 3-clique in the newly added city except for O$_3$. This additional city surrounds the existing 3-clique geographically, setting the PM$_{2.5}$ for an example, the PMFG filters the Taicang, Kunshan and Shanghai as the highest strong 3-clique, and the added Changshu in 4-clique is very close to the previous three cities. These cities are the manufacturing centers and energy-intensive centers of the Yangtz River Delta. Moreover, the amount of private cars ranks high in China, resulting in a strongly correlated clique in terms of the most localized pollutants. The 4-clique for O$_3$ has been switching from Hunan Province to Liaoning Province rather than append another city on the basis of 3-clique, showing the unstable structure of O$_3$'s 3-clique, One possibility is that Liaoning Province has more industrial companies to accommodate more individuals when the correlated community expands.  In Table~\ref{Tab5}, we also tabulate each strongest clique's averaged correlations (measured by the average of 3 pair-wise correlations in 3-clique and 6 pair-wise correlations in 4-clique) and the extension from 3-clique to 4-clique reduces the average correlation by about 0.02 for each pollutant. The last column displays the averaged disparity measure $y_{i}$:
\begin{equation}
y_i=\sum_{j\neq i, j\in{\mathrm{clique}}} \left[\frac{c_{ij}}{s_i}\right]^2
\label{Eq:PMFG:Disparity:Def}
\end{equation}
where $s_i=\sum_{j\neq i, j\in clique} c_{ij}$. If the correlation is uniform across each intercity pair within the clique, 3-clique's $y_{i}=1/2$ and 4-clique's $y_{i}=1/3$. The last column shows an overall uniform correlations within each strongest correlated clique.

In this section, we resort to the administrative divisions as a rough measure of the cities' geographic distance, although the results that most highest correlated cliques are centered within one or two provinces are pretty straightforward. Critics may point out that two cities in different provinces are even closer than two cities in the same province. In an ongoing research, we are quantitatively measuring the spatial correlated structures with their mutual distances.

\begin{table*}
\centering
\caption{{\bf Strongest correlated clique. This table reports the strongest correlated 3-clique or 4-clique for each pollutant.}}
\label{Tab5}
\setlength\tabcolsep{1pt}
\tiny
\begin{tabular}{ccccccc}
\hline
Pollutant &City 1 & City 2 &City 3 &City 4 &$\langle c \rangle$&$\langle y \rangle$ \\
\hline
\multicolumn{7}{c}{Panel A: 3-clique strongest correlated clique }\\
\hline
 PM$_{2.5}$&  Taicang (Jiangsu) & Kunshan (Jiangsu) & Shanghai (Shanghai) && 0.92 & 0.5000\\
 PM$_{10}$&   Changzhou (Jiangsu) & Wuxi (Jiangsu) & Jiangyin (Jiangsu) && 0.90& 0.5002 \\
 CO&   Suzhou (Jiangsu) & Shanghai (Shanghai) & Jiaxing (Zhejiang) && 0.80 & 0.5004\\
  O$_3$&  Zhuzhou (Hunan) & Xiangtan (Hunan) & Changsha (Hunan) && 0.92& 0.5011 \\
 NO$_2$&   Xiangtan (Hunan) & Hengyang (Hunan) & Changsha (Hunan) && 0.82& 0.5001 \\
 SO$_2$&   Harbin (Heilongjiang) & Changchun (Jilin) & Shenyang (Liaoning) && 0.82& 0.5000\\
\hline
\multicolumn{7}{c}{Panel B: 4-clique strongest correlated clique }\\
\hline
PM$_{2.5}$&    Taicang (Jiangsu) & Changshu (Jiangsu) & Kunshan (Jiangsu) & Shanghai (Shanghai) & 0.90  & 0.3335 \\
 PM$_{10}$&   Changzhou (Jiangsu) & Zhangjiagang (Jiangsu) & Wuxi (Jiangsu) & Jiangyin (Jiangsu) & 0.88  & 0.3337 \\
CO&    Jimo (Shandong) & Jiaozhou (Shandong) & Laixi (Shandong) & Qingdao (Shandong) & 0.79  & 0.3338 \\
   O$_3$&   Benxi (Liaoning) & Shenyang (Liaoning) & Liaoyang (Liaoning) & Anshan (Liaoning) & 0.89  & 0.3340 \\
   NO$_2$&   Zhuzhou (Hunan) & Xiangtan (Hunan) & Hengyang (Hunan) & Changsha (Hunan) & 0.80  & 0.3336 \\
    SO$_2$&  Harbin (Heilongjiang) & Changchun (Jilin) & Benxi (Liaoning) & Shenyang (Liaoning) & 0.80  & 0.3339 \\
\hline
\end{tabular}
\end{table*}

\subsection*{Cliques' wan and wax}

\begin{figure}
  \includegraphics[width=0.95\linewidth]{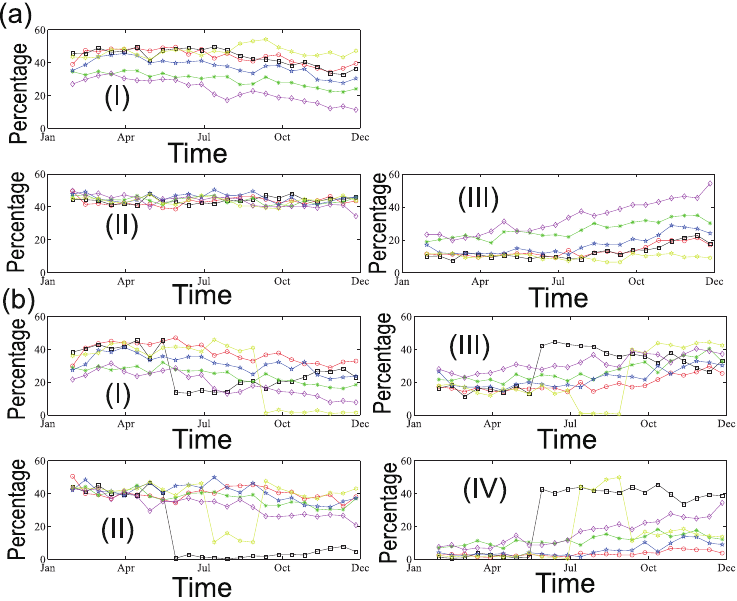}
  \caption{{\bf The time evolution of the percentages that all the cities in a clique belong to one province, two provinces, three or four provinces. (a) is 3-clique's evolutionary results with (I)-(III) corresponding to one province, two provinces and three provinces respectively,  (b) is 4-clique's evolutionary results with (I)-(IV) corresponding to one province, two provinces, three provinces and four provinces respectively. Legend of each pollutant refers to Fig.~\ref{Fig2}(b) or (c).}}
\label{Fig7}
\end{figure}

The moving window scheme from Eq.~(\ref{Eq:MovingWindows:CorrelationDefinition}) shows how the percentages of the nodes in a clique belonging to one province, two provinces, three and four provinces evolve all over the whole 2015. Fig.~\ref{Fig7}(a) show the 3-clique dynamic patterns. Among all the cliques, except the O$_3$, the other five air pollutants' one province 3-clique percentages commonly have a declining trend, which acts as a strong signal for the anti-localization and diffusion of the pollutants. However, the percentages of cliques that belong to two different provinces are rather stable varying from 40\% to 50\%. And the reduced part in one province percentages flows into the three provinces percentages. Especially for the SO$_2$, the higher correlated 3-clique is more pervasive in three different provinces, and up to 60\% of all 3-cliques at the end of 2015. On the other hand, the localization of O$_3$ is quite straightforward: 45\% of 3-cliques are in one province and another 45\% are in two different provinces, leaving only 10\% dispersion in three different provinces. When we extend the 3-clique to 4-clique in Fig.~\ref{Fig7}(b), the scenario is quite different. O$_3$'s localization is not stable any more and the mutation period occurs around September. Before September, the percentage of 4-clique whose cities are in the same provinces stabilizes at around 40\%. However, after September the 4-clique structures become more diversified. Another evident breakpoint happens to PM$_{10}$ around late May. Before May, most of the 4-cliques pertain to same province or two provinces, indicating the strong local effect of PM$_{10}$. However,  after May this tide has been reversed to the extent in which three provinces and four provinces individuals dominate the 4-clique. The rest four pollutants are slightly decreasing the local tendency and increasing the diverging correlations. How climatic conditions and human activities influence the divergent trend needs to be evaluated further. One thing is for sure that the two breakpoints are both climate change points and industrial activity peak time in China \cite{Cao-Lee-Ho-Zou-Fung-Li-Waston-Chow-2004-AtmosEnv}. To sum up, we are fine to conclude that the correlation structure of the pollutants are in a course of slightly divergent dispersion in space. This finding, to some extent, consolidates the hypothesis that Chinese air pollution's diffusion power is further reaching.

\begin{figure}
  \includegraphics[width=0.95\linewidth]{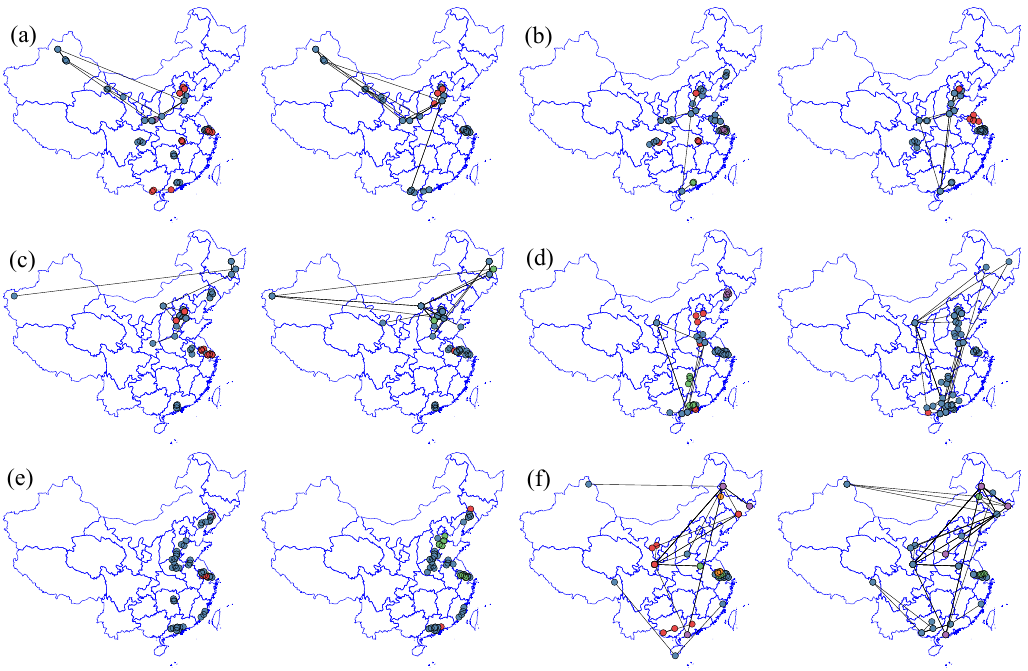}
  \caption{{\bf The residence time for the most strongly correlated PMFG filtered 3-cliques and 4-cliques. The filled circles in different colors (blue, red, green, orange, purple) are 5 equally increased segments of the number of days the clique lasts throughout the year. Left panel is the six pollutants graphed from the most strongly correlated 3-cliques and right panel is based on the most strongly correlated 4-cliques. In each plot, from (a) to (f) are PM$_{2.5}$, PM$_{10}$, CO, NO$_2$, O$_3$ and SO$_2$ respectively.}}
\label{Fig8}
\end{figure}

\begin{figure}[ht]
  \includegraphics[width=0.7\linewidth]{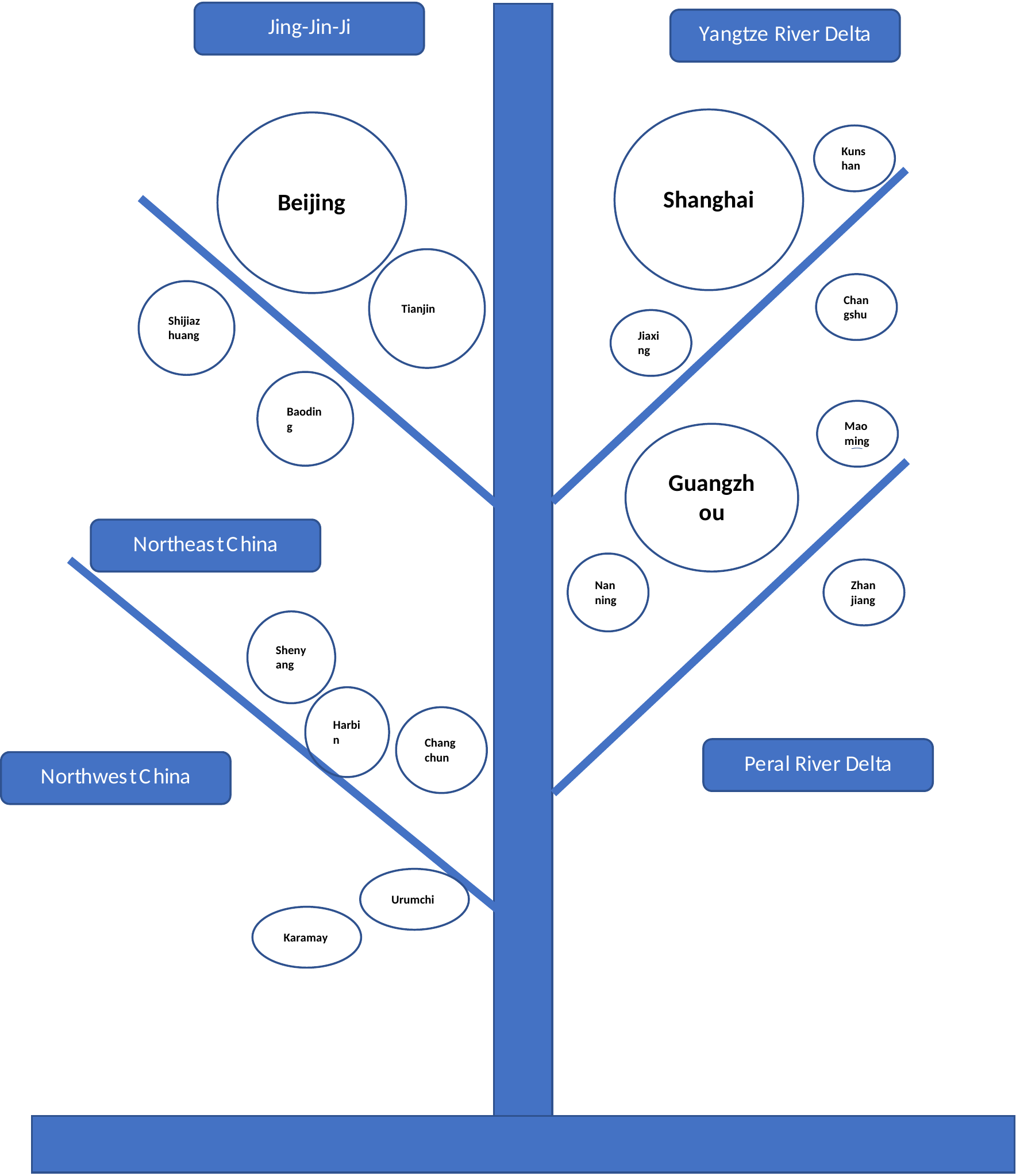}
  \caption{\bf{Cities' evolution tree based on the dynamic strongest correlated cliques. After plotting the frequency of the strongest correlated cliques in Fig.~\ref{Fig8}, we classify the strongest correlation structure into six parts. The area of circle roughly represents the population, and the cities in the top of tree have higher GDP. Each branch is a city group.}}
\label{Fig9}
\end{figure}

Consisting with the previous static analysis, we plot the evolutions of the most strongly correlated cliques across each sliding window and find that even though different strongest cliques will occur in different windows, only limit to the several individual cities labeled in Fig.~\ref{Fig8}. By and large, both Shanghai-centered and Beijing-centered city groups are ubiquitous in each pollutant's cliques. Although the two cities are service-oriented economy, this result is not surprising because there are thousands of state-run and private-owned manufacturing industries distributed along the two economic belts and the density of population is among the highest as shown in Fig.~\ref{Fig9}, resulting in highly correlated spatial community of air pollutants in these cities. On the other hand, the policy implication is straightforward, collaborative management and joint monitoring will be more effective in controlling the pollution in these two areas. The 4-clique generally has boarder coverage than 3-clique and the connections labeled as solid lines are denser. Moreover, the length of these edges connecting cities signals the exterior extension degree. In this sense, O$_3$'s strongest cliques display visibly localized properties, while strongest correlated cliques of CO and SO$_2$ even connect extremely western and eastern cities of China. The strongest correlated cliques' occurrence probability (measured by the number of clique's lasting days and labeled with five colors) reveals that the above-mentioned two city groups are more likely to be encompassed into these highly correlated cliques.

To gross all the stable strongest correlated cliques using cities evolution tree \cite{Wang-Liu-Peng-Chen-Driskell-Zheng-2012-PE}, we find in Fig.~\ref{Fig9} that five parts of China are of great significance in understanding the dynamic spatial structure. The Beijing-centered Jing-Jin-Ji belt and the Shanghai-centered Yangtze-River Delta as well as the Guangzhou-centered Pearl-River Delta are strongly correlated because of distributed intensive manufacturing firms \cite{Chan-Yao-2008-AtmosEnv}. The cities in Northeast China and Northwest China are strongly correlated, which is mainly caused by the large amount of pollutants emitted due to the heating supply in winter \cite{Tao-2016-AtmosEnv}. Moreover, particulate pollution in these parts is under the common influence of regionally accumulated pollutants due to the lack of strong winds.

In brief, the dynamic analysis above allows us to check the stability of both these cliques themselves and their relationship with geographical distances. Except one or two particular pollutants, PMFG filters these correlated cliques which is characterized by an overall stable but slightly divergent trend in space. In addition, the most strongly correlated cliques apparently center around some developed areas for rather a long period. These findings will encourage us to think about a cooperative management model to curb the localized cliques and decentralized method to control divergent pollutants such as SO$_2$.

\section*{Discussion}

Chinese air pollution arises rapidly along with the economic development and is torturing the whole country\cite{Chan-Yao-2008-AtmosEnv}. It requires tenacious determination to fight against the air pollution. This paper serves to deepen our understanding of the air pollutants evolutionary process from temporal and spatial correlations.

To begin with, we depict the intraday patterns for five pollutants in terms of two contraction periods and two expansion periods. O$_3$, as an exception, shows its asynchronism. Moreover, this trend is pervasive all over Chinese cities and across all four seasons. The Lomb spectrum analysis proves that these intraday patterns are daily periodical across the whole year. These results help us overview the general structure of air pollutants' time series and lay foundations for the understanding of temporal and spatial correlations.

From the temporal side, most cities' air pollutants' time series show strong long-term correlations, which is consistent with the trend that smoggy days are always followed by one another. This finding can be partly explained by the successive random dilution model \cite{Ott-1994}, where the air pollutants undergo a random dilution and mixing process and accumulate again, resulting in a multiplicative dilution process \cite{Lee-2002-WASP}. To further explore the spatial heterogeneity of these Hurst exponents stratified by socioeconomic and geographical indicators \cite{Wang-Haining-Liu-Li-Jiang-2013-EPA,Wang-Zhang-Fu-2016-EcolInd}, we find that the Hurst exponents are more heterogenous partitioned by the geographical indicator, which shows that the long-term structures of air conditions are closer in cities at similar development levels. This finding also partially shows that human factors outweigh natural factors in determining the long-term trend of air pollution \cite{Lee-2002-WASP}.  We also find two particulate matters share the similar temporal trends; other three pollutants (CO, NO$_2$ and SO$_2$) also behave similarly in the long-term correlations. This particularity of O$_3$ is largely due to its asynchronous changing process with other pollutants.  The relationship between Hurst exponents and several basic statistics is also displayed, although the results are mixed, we capture a negative correlation between $H$ and skewness or kurtosis.

The policy implication of the long-term structure is twofold. First, except for O$_3$, the other five pollutants' long-term correlations inform us that weakening the multiplicative dilution and accumulation process of air pollutants requires a comprehensive set of actions based on an integrated approach to make substantial improvements \cite{Bhardwaj-Pruthi-2016-NGCT}. Second, assessing the spatial stratified heterogeneity, the relationship between Hurst exponents and other statistics makes regional variations of pollutants' long-term structure clear, providing an empirical support in the prediction of pollutants' evolution \cite{Shen-Huang-Yan-2016-PA}.

On the spatial side, starting from the Pearson correlation structure, we are the first to cover almost every medium-sized Chinese cities to unveil the spatial correlations compared to previous researches \cite{Li-Qian-Qu-Zhou-Guo-Guo-2014-EnvPol,Wang-Ying-Hu-Zhang-2014-EnvirInter}. It's shown that O$_3$ tops the six pollutants in terms of overall correlations. All the six correlation spectrum are featured with clusters. We then successfully refine small cliques in the correlation structure aided by PMFG \cite{Tumminello-Aste-DiMatteo-Mantegna-2005-PNAS}.
These cliques reflect how the spatial similarity of the pollutants' evolutionary process looks like and show how the six pollutants disperse spatially. We confirm that neighbouring cities are more likely to form clusters \cite{Chan-Yao-2008-AtmosEnv}. Based on the spatial correlations of each pollutant, we classify the pollutants into three groups with increasing dispersion power: Particulate matters (PM$_{2.5}$ and PM$_{10}$), O$_3$ and NO$_2$ which merely spread to the third or forth provinces, SO$_2$ and CO which are easy to form cliques with cities far away. The tabulated highest correlated cliques show that manufacturing centers are more likely to form strong correlation structure \cite{Chan-Yao-2008-AtmosEnv,Tao-2016-AtmosEnv,Tumminello-Lillo-Mantegna-2010-JEBO}. These well-identified small cliques are of great value in understanding the pollution's spatial correlations\cite{Tumminello-Aste-DiMatteo-Mantegna-2005-PNAS,Tumminello-DiMatteo-Aste-Mantegna-2007-EPJB}.

Finally, we test the correlation's dynamic stability across the year through a moving window scheme. It is found that O$_3$ has breakpoints in both 3-clique and 4-clique around September, and PM$_{10}$ also shows its breakpoint around late May, while other pollutant present a general stable divergent and diffusive trend in spatiality. These two breakpoints can be partly explained by climate change points and industrial activity peak times in China \cite{Cao-Lee-Ho-Zou-Fung-Li-Waston-Chow-2004-AtmosEnv}. The finding that the correlation structure of pollutants is slightly divergent serves a piece of solid evidence that air pollution in China is reaching further away, making the environmental issue severer \cite{Chan-Yao-2008-AtmosEnv,Tao-2016-AtmosEnv}.

Although these conclusions are carefully drawn and cautiously presented, we still have huge potential to improve. First, the causes of a decaying correlation between two cities are rather complicated, with the distance being one of the many factors. Other meteorological conditions such as wind largely account for this spatial correlation patterns. Modelling this spatial correlation with distance, possibly leads to a unilateral conclusive result. In the future study, we should try to collect other meteorological data to enhance the causes of spatial connections. Second, although we unveil the pollutants' spatial patterns from both static and dynamic analyses, the shifting patterns remain a puzzle. To achieve this end, two difficulties are ahead. The first is to introduce a diffusion model among so many cities and the second is to identify the correlation directly coming from pollutants' shifting rather than data noise. Anyway, it provides a promising direction for future research.

\section*{Acknowledgement}

We appreciate Shanghai Qingyue Open Environmental Protection Data Center (QOEPDC) for providing the hourly data and Miss Faping Yang (Social and Public Administration School in East China University of Science and Technology) collected all the Chinese cities' information. This work is partially supported by the China Scholarship Council (20150674) and the Fundamental Research Funds for the Central Universities (222201718006).


%
%
%


\end{document}